\documentclass[prd,nofootinbib,showpacs,12pt]{revtex4}

\usepackage{graphicx}

\begin{document}

\title{Towards Field Theory in Spaces with Multivolume Junctions}

\author{P.~I.~Fomin\thanks{E-mail: pfomin@bitp.kiev.ua}
and Yu.~V.~Shtanov\thanks{E-mail: shtanov@bitp.kiev.ua}} \affiliation{Bogolyubov
Institute for Theoretical Physics, Kiev 03680, Ukraine}

\date{January 10, 2002}

\begin{abstract}
We consider a spacetime formed by several pieces having common timelike boundary which
plays the role of a junction between them.  We establish junction conditions for fields
of various spin and derive the resulting laws of wave propagation through the junction,
which turn out to be quite similar for fields of all spins. As an application, we
consider the case of multivolume junctions in four-dimensional spacetime that may arise
in the context of the theory of quantum creation of a closed universe on the background
of a big mother universe.  The theory developed can also be applied to braneworld models
and to the superstring theory.
\end{abstract}


\maketitle

\section{Introduction}

\begin{figure}
\begin{center}
\includegraphics[width=0.5\textwidth]{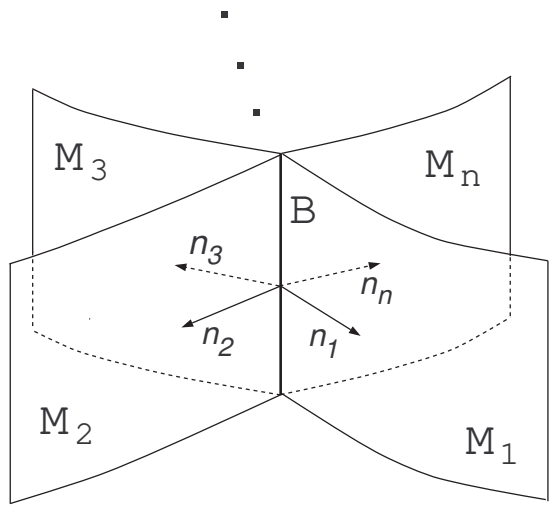}
\end{center}
\caption{\small Multivolume junction.\label{space-n}}
\end{figure}

\begin{figure}
\begin{center}
\includegraphics[width=0.58\textwidth]{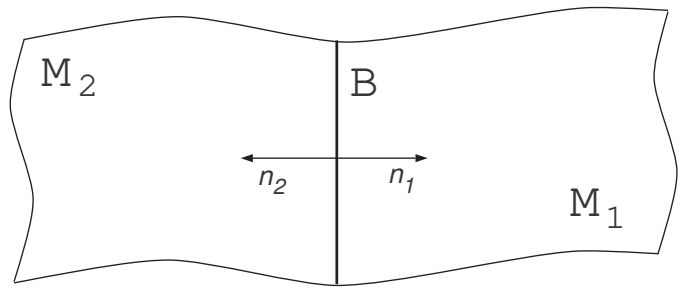}
\end{center}
\caption{\small Two-volume junction.\label{space-2}}
\end{figure}

In this paper, we consider field theory on spaces with the so-called multivolume
junctions, as shown in Fig.~\ref{space-n}.  Spaces of such topological configurations
arise in several important contexts. Firstly, they arise in the study of various
braneworld theories that became very popular after the seminal papers \cite{RS}. In these
theories, the dimensionality of the volume spaces is usually equal to five, and the
junction, which is called brane, is four-dimensional and is identified with the physical
spacetime. Normally, the brane in a braneworld theory separates two volume spaces; the
corresponding situation is shown in Fig.~\ref{space-2}. However, the brane may be a
junction of more than two volume spaces as well as it may be a boundary of only one
volume space. Thus, in \cite{CS}, junctions of an arbitrary number of semi-infinite
four-branes were under consideration, and the whole configuration was assumed to be
embedded in a six-dimensional spacetime. Secondly, such spaces may represent the
situation where the physical four-dimensional spacetime has a nontrivial topology of the
type shown in Fig.~\ref{space-n}. Thirdly, in the important case of two-dimensional
volume spaces, our investigation may be applicable to the superstring theory. Note that
three-volume junctions of boson strings of type shown in Fig.~\ref{space-n} were under
consideration in \cite{string}.

In this paper, we consider the general situation depicted in
Fig.~\ref{space-n}. It symbolically shows $n$ $d$-dimensional Lorentzian
manifolds ${\mathcal M}_s$, $s = 1, \ldots, n$, with common \mbox{$(d\!-\!
1)$}-dimensional boundary ${\mathcal B}$ which thus plays the role of a
junction between them.  The boundary ${\mathcal B}$ is assumed to be
timelike, so that all the respective inner normal vector fields $n^a_{(s)}$,
$s = 1, \ldots, n$, at this boundary are spacelike.  We do not consider the
space of Fig.~\ref{space-n} as embedded in a higher-dimensional manifold.
Our aim is to study the behavior of various fields in a space with the
specified topology focussing attention on the derivation and analysis of the
junction conditions at ${\mathcal B}$.

In brane theories, together with fields in the volume, one also considers
fields whose dynamics is restricted to the brane \cite{RS}. Moreover, the
action for the brane may involve the restrictions of some of the volume
fields to the brane; for example, it typically involves the induced metric.
However, in this paper, we restrict ourselves mainly to the case where the
junction ${\mathcal B}$ does not have its intrinsic dynamics and thus
represents what might be called a generalization of an imaginary
hypersurface separating two volumes in an ordinary space to the case where
the number of volumes is greater than two.

\begin{figure}
\begin{center}
\includegraphics[width=0.55\textwidth]{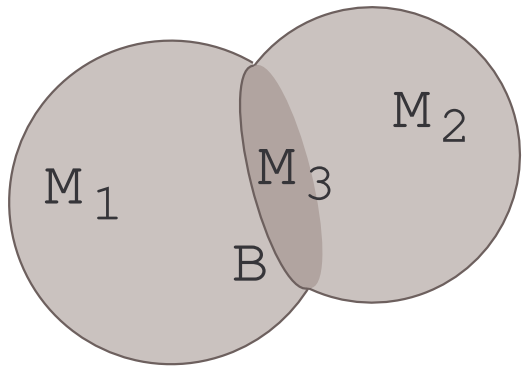}
\end{center}
\caption{\small Closed universe with multivolume junctions.\label{branch}}
\end{figure}

As a concrete example of application of our theory, we consider the case of
multivolume junctions in a four-dimensional spacetime
(Sec.~\ref{branching}). The issue of such a junction may arise in the
context of the theory of quantum creation of a closed universe on the
background of a big mother universe \cite{Fomin}. It is conceivable that the
created baby universe does not become spatially separated from the mother
universe, but rather remains glued with it over some common
three-dimensional volume \cite{FSB}.  The corresponding situation is
depicted in Fig.~\ref{branch}, which shows the mother universe ${\mathcal
M}_1$ and the baby universe ${\mathcal M}_2$ glued over the volume
${\mathcal M}_3$.  All the three volume regions ${\mathcal M}_1$, ${\mathcal
M}_2$, and ${\mathcal M}_3$ may evolve metrically preserving the topological
configuration as shown in Fig.~\ref{branch}.  One of the important physical
questions in this situation is the issue of the behavior of various physical
fields in this topology, in particular, the conditions of propagation of
waves through the junction ${\mathcal B}$ which is the common boundary of
all three volume regions.

In approaching this issue, we first establish junction conditions for fields
of various spins (Secs.~\ref{metric}--\ref{spin}) and then consider the
resulting laws of wave propagation through the junction ${\mathcal B}$
(Sec.~\ref{wave}). In principle, the junction conditions at ${\mathcal B}$
may be specified in many different ways. However, with a natural requirement
that the spaces ${\mathcal M}_s$ be treated identically, it turns out that
there are precisely two versions of junction conditions for each spin. This
is one of the reasons why we pay attention to spaces with the topology
specified above and propose to study them in greater detail.

In Sec.~\ref{green}, we study the Green functions in the case of junction of
flat spaces over a flat boundary and the resulting quantum vacuum
stress-energy tensor.

Concluding remarks are presented in Sec.~\ref{discussion}.

\section{Junction conditions for the metric} \label{metric}

We start with considering junction conditions for the metric because the
actions for all other fields involve ingredients (for example, the volume
element) associated with the metric.  In this paper, we consider the metric
$g_{ab}$ with signature $+ - - \ldots -$. The general action for the metric
can be written in the form \cite{GH}
\begin{equation}
S =  - M^{d - 2} \left(\int_{\mathcal M} R + 2  \sum_s \int_{\mathcal B}
K^{(s)} \right) + \int_{\mathcal B} L_h\, , \label{metricac}
\end{equation}
where $M$ is the Planck mass, $R$ is the scalar curvature, and $K^{(s)}$ is
the trace of the extrinsic curvature of the junction ${\mathcal B}$ in the
space ${\mathcal M}_s$. We impose the most natural junction conditions for
the metric, namely, that the induced metric $h_{ab}$ on ${\mathcal B}$ is
one and the same in all the spaces ${\mathcal M}_s$, $s = 1, \ldots, n$. The
Lagrangian $L_h$ in (\ref{metricac}) depends only on this induced metric.

In this paper, we use the notation and conventions of \cite{Wald}.  The
extrinsic curvature and its trace are defined as follows:
\begin{equation}
K_{ab} = h^c{}_a \nabla_c n_b \, , \quad K = K_{ab} h^{ab} \, ,
\end{equation}
where $h_{ab} = g_{ab} + n_a n_b$ is the metric induced on a timelike
hypersurface. The natural volume elements are implied in all the
integrations over ${\mathcal M}$ and ${\mathcal B}$.  The
cosmological-constant terms can be added to action (\ref{metricac}) for each
space ${\mathcal M}_s$, and their contribution to the resulting equations is
obvious.

Variation of action (\ref{metricac}) can be written in the form (see
Appendix)
\begin{equation}
\delta S = - M^{d - 2} \left[\int_{\mathcal M} G_{ab} \delta g^{ab} +
\int_{\mathcal B} \sum_s \left( K_{ab}^{(s)} - K^{(s)} h_{ab} \right) \delta
h^{ab} \right] + \int_{\mathcal B} \sigma_{ab} \delta h^{ab}\, .
\label{metricvar}
\end{equation}
Here, $G_{ab}$ is the Einstein tensor, $\sigma_{ab}$ is the variation of the
last term in (\ref{metricac}) with respect to $h^{ab}$, and the variation
$\delta h^{ab}$ is completely determined by the variation $\delta g^{ab}$ of
the metric in ${\mathcal M}$. Note that the Gibbons--Hawking boundary terms
\cite{GH} in action (\ref{metricac}) are required to consistently obtain the
Einstein equations in the respective volume spaces without restricting
variations of the metric at the junction ${\mathcal B}$.

Besides the metric, we may have additional fields $\Phi$ that propagate in
the volume and fields $\varphi$ whose dynamics is restricted to the
junction. Some of the fields $\varphi$ may represent restrictions of some of
the volume fields $\Phi$ to the junction. Thus, in general, we must consider
the action for the fields with the corresponding Lagrangians $L_\Phi$ and
$L_\varphi$ in the form
\begin{equation}\label{matter}
\int_{\mathcal M} L_\Phi + \int_{\mathcal B} L_\varphi \, .
\end{equation}

The additional set of junction conditions obtained with the account of
(\ref{metricvar}), (\ref{matter}) is then
\begin{equation}
\sum_s \left(K_{ab}^{(s)} - K^{(s)} h_{ab} \right) = {1 \over M^{d-2}}
\left(\sigma_{ab} + \tau_{ab} \right) \, , \label{metric1}
\end{equation}
where $\tau_{ab}$ is the variation of (\ref{matter}) with respect to the
induced metric $h^{ab}$.

Variation of the first term in (\ref{matter}), together with
(\ref{metricvar}), leads to the Einstein equation in the volume
\begin{equation}
G_{ab} = {1 \over M^{d-2}}\, T_{ab} \, ,
\end {equation}
where $T_{ab}$ is the variation of the first term in (\ref{matter}) with
respect to the metric $g^{ab}$.

\section{Junction conditions for scalar and vector fields} \label{scalvec}

A complex scalar field $\phi$ with mass $m$ is described by the action
\begin{equation}
S = \int_{\mathcal M} \left[\overline{\nabla_a \phi} \nabla^a\phi - m^2
\bar\phi\phi \right] \, , \label{scalac}
\end{equation}
where the bar denotes complex conjugation, the integral is taken over the
whole manifold ${\mathcal M} = \cup_s {\mathcal M}_s$ shown in
Fig.~\ref{space-n}, and the natural volume element is implied. The
derivative $\nabla_a \phi$ may involve contribution from the gauge vector
field.

In formulating the junction conditions at ${\mathcal B}$, we proceed from
the following natural principles.  Let $\phi_{(s)}$ denote the restriction
of the scalar field $\phi$ to the space ${\mathcal M}_s$.  We are going to
relate the value of $\phi_{(s)}$ with the values of $\phi_{(r)}$, $r \ne s$
at the junction. This relation must be linear (in order to respect the
superposition principle), and the spaces ${\mathcal M}_s$ must be regarded
as physically identical. With these requirements, we arrive at the following
general junction conditions at ${\mathcal B}$:
\begin{equation}
\phi_{(s)} = \alpha \sum_{r \ne s} \phi_{(r)} \, , \quad s = 1, \ldots, n \,
, \label{scalbc}
\end{equation}
where $\alpha$ is some constant to be determined. Possible values of
$\alpha$ are obtained from the additional requirement that the junction
conditions (\ref{scalbc}) allow for nontrivial solutions at the junction.
This gives only two possible values of the parameter $\alpha$:
\begin{equation}
\alpha = {1 \over n -1} \quad \mbox{and} \quad \alpha = -1 \, .
\end{equation}
Notably, the condition $\alpha = 1/(n - 1)$ simply implies the {\em
continuity\/} of the scalar field in the space ${\mathcal M} = \cup_s
{\mathcal M}_s$, i.e., the condition $\phi_{(1)} = \phi_{(2)} = \cdots =
\phi_{(n)}$, while the condition $\alpha = -1$ leads to the single equation
$\sum_s \phi_{(s)} = 0$. To obtain other conditions at the junction, we vary
the action respecting the junction conditions (\ref{scalbc}) and demanding
that the variation be zero. General variation of action (\ref{scalac}) is
given by
\begin{equation}\begin{array}{ll}
\delta S = &\displaystyle - \int_{\mathcal M} \left[\delta \bar \phi
\left(\nabla^a \nabla_a + m^2 \right) \phi + \delta \phi \left(\bar \nabla^a
\bar \nabla_a + m^2 \right) \bar \phi\right] \medskip \\ &\displaystyle +
\sum_s \int_{\mathcal B} \left( \delta \bar \phi_{(s)} \nabla_a \phi_{(s)} +
\delta \phi_{(s)} \bar \nabla_a \bar \phi_{(s)} \right) n^a_{(s)} \, .
\label{scalvar}
\end{array}
\end{equation}
According to the value of $\alpha$ in (\ref{scalbc}), we obtain, besides the
Klein--Gordon equations of motion in the volume, also the additional
junction conditions.

\subsection*{A. Case of $\alpha = 1/(n -1)$}

In this case, the additional junction conditions are $\sum_s n^a_{(s)}
\nabla_a \phi_{(s)} = 0$. We summarize the junction conditions obtained in
the case under consideration:
\begin{equation}
\phi_{(p)} = \phi_{(q)} \quad \mbox{for all $p, q$}\, ; \quad \sum_s
n^a_{(s)} \nabla_a \phi_{(s)} = 0 \, . \label{scal1}
\end{equation}

\subsection*{B. Case of $\alpha = -1$}

In this case, the complete set of junction conditions is
\begin{equation}
\sum_s \phi_{(s)} = 0 \, ; \quad n^a_{(p)} \nabla_a \phi_{(p)} = n^a_{(q)}
\nabla_a \phi_{(q)} \quad \mbox{for all $p, q$}\, . \label{scal2}
\end{equation}

Both sets of junction conditions (\ref{scal1}) and (\ref{scal2}) imply the
sum rule for the components of the conserved current $J_a = i \left(\bar
\phi \nabla_a \phi - \phi \bar \nabla_a \bar \phi \right)$ normal to the
junction surface ${\mathcal B}$:
\begin{equation}
\sum_s n^a_{(s)} J_a^{(s)} = 0 \, . \label{current}
\end{equation}
The junction conditions (\ref{scal1}) also imply that the components of the
current along the junction surface ${\mathcal B}$ are the same in all the
volume spaces.

A free vector field $A^a$ with mass $m$ is described by the action
\begin{equation}
S = \int_{\mathcal M} \left(- \frac14 F_{ab} F^{ab} + {m^2 \over 2} A_a A^a
\right) \, . \label{vecaction}
\end{equation}
Its variation in the manifold ${\mathcal M}$ is given by
\begin{equation}
\delta S = - \int_{\mathcal M} \delta A_b \left( \nabla_a F^{ab} + m^2 A^b
\right) - \sum_s \int_{\mathcal B} \delta A_b^{(s)} F^{ab}_{(s)} n_a^{(s)}
\, . \label{vecvar}
\end{equation}

The junction conditions at ${\mathcal B}$ naturally will involve only the
components of $A^a$ tangent to ${\mathcal B}$, which we denote by
$A^{\parallel a}$. Similarly to Eq.~(\ref{scalbc}), we write the most
general expression compatible with the symmetry imposed:
\begin{equation}
A^{\parallel a}_{(s)} = \alpha \sum_{r \ne s} A^{\parallel a}_{(r)} \, , \ \
s = 1, \ldots, n \, ,  \label{vecbc}
\end{equation}
where, by reasoning similar to that of the scalar case, the constant
$\alpha$ is equal either to $1/(n - 1)$ or to $-1$.

\subsection*{A. Case of $\alpha = 1/(n - 1)$}

In this case, the values of $A^{\parallel a}$ are the same at all the sides
of the junction ${\mathcal B}$.  Variation of action (\ref{vecvar}) yields
the Proca equations (the Maxwell equations, if $m = 0$) in the volume and
the additional junction conditions $\sum_s F^{ab}_{(s)} n^{(s)}_b = 0$.  We
summarize the junction conditions in this case:
\begin{equation}
A^{\parallel a}_{(p)} = A^{\parallel a}_{(q)} \quad \mbox{for all $p, q$}\,
; \quad \sum_s F^{ab}_{(s)} n^{(s)}_b = 0 \, . \label{vec1}
\end{equation}

\subsection*{B. Case of $\alpha = -1$}

In this case, the total set of junction conditions obtained with the account
of (\ref{vecvar}) is given by
\begin{equation}
\sum_s A^{\parallel a}_{(s)} = 0 \, ; \quad  F^{ab}_{(p)} n^{(p)}_b =
F^{ab}_{(q)} n^{(q)}_b \quad \mbox{for all $p, q$}\, . \label{vec2}
\end{equation}

In the case $n = 1$, i.e., where there is only one volume space with
boundary, the junction conditions of type A [Eqs.~(\ref{scal1}),
(\ref{vec1})] become the Neumann boundary conditions, while the junction
conditions of type B [Eqs.~(\ref{scal2}), (\ref{vec2})] become the Dirichlet
boundary conditions.  Thus, the junction conditions obtained above can be
regarded as respective generalizations of the mentioned boundary conditions
to the case of $n > 1$.

\section{Junction conditions for the Dirac field} \label{spin}

The action for the Dirac field of mass $m$ can be written in the form
\begin{equation}
S =  \int_{\mathcal M} \left( {i \over 2} \left[ \overline \psi \gamma^a
\nabla_a \psi - \overline{(\nabla_a \psi)} \gamma^a \psi \right] - m
\overline \psi \psi \right) \, , \label{spinac}
\end{equation}
where $\nabla_a \psi$ denotes the standard covariant derivative of the Dirac
spinor $\psi$ that may include contribution from the gauge vector field.
Variation of action (\ref{spinac}) is given by
\begin{equation}\begin{array}{ll}
\delta S = &\displaystyle \int_{\mathcal M} \delta \overline \psi \left(i
\gamma^a \nabla_a - m \right) \psi - \int_{\mathcal M} \left[ i
\overline{(\nabla_a \psi)} \gamma^a + m \overline \psi \right] \delta \psi
\medskip \\ &\displaystyle {} + i \sum_s \int_{\mathcal B} \left(\overline
\psi_{(s)} \gamma^a \delta \psi_{(s)} - \delta \overline \psi_{(s)} \gamma^a
\psi_{(s)} \right) n_a^{(s)} \, , \label{spinvar}
\end{array}
\end{equation}
where $\psi_{(s)}$, $\overline \psi_{(s)}$ are the restrictions of the
spinor fields to the space ${\mathcal M}_s$.

The specific feature of the spinor field is that it is referred at each
point to a particular orthonormal basis (called tetrad in the case of $d =
4$). Thus, in order to formulate the junction conditions for the Dirac
field, we must take this circumstance into account. We need to relate the
spinor fields $\psi_{(s)}$, $s = 1,\ldots,n$, as we reach one and the same
point at the junction ${\mathcal B}$ moving in different spaces ${\mathcal
M}_s$. At each point $x \in {\mathcal B}$, we must choose $n$ orthonormal
bases, one in each space ${\mathcal M}_s$, $s = 1,\ldots,n$, to refer the
corresponding values of the Dirac field to these bases.  For a convenient
formulation of the relations between these values, the $n$ bases are to be
chosen in a coherent way.  We choose $d - 1$ of the basis vectors $\{e_1^a,
e_2^a, \ldots, e_{d-1}^a\}$ at any point $x \in {\mathcal B}$ to form an
arbitrary orthonormal basis in the tangent space to ${\mathcal B}$, the same
for all the spaces ${\mathcal M}_s$, $s = 1,\ldots,n$. Then the $d$-th
vector of the orthonormal basis in each space is determined uniquely up to
sign, and we choose it to be the inner normal vector $n_{(s)}^a$, $s =
1,\ldots,n$, respectively, in each of these spaces.

Next, to obtain the possible junction conditions at ${\mathcal B}$, we first
consider the case $n = 2$, i.e., a manifold ${\mathcal M}$ and a timelike
hypersurface ${\mathcal B}$ that divides it into two parts ${\mathcal M}_1$
and ${\mathcal M}_2$ (see Fig.~\ref{space-2}). Given a Dirac spinor field
which is {\em continuous\/} in ${\mathcal M}$, its values at ${\mathcal B}$
in the bases $\{e_1^a, e_2^a, \ldots, e_{d-1}^a, n_{(s)}^a\}$, $s = 1,2$,
chosen by the procedure described above are related by the operation of
reflection which performs the transformation of the Dirac spinor $\psi$ from
the basis $\{e_1^a, e_2^a, \ldots, e_{d-1}^a, n_{(s)}^a\}$ to the reflected
basis $\{e_1^a, e_2^a, \ldots, e_{d-1}^a$, $- n_{(s)}^a\}$ (since $n_2^a = -
n_1^a$).  For simplicity, we consider the important case where the
dimensionality $d$ is even and define the corresponding matrix operator of
reflection as $N = \gamma_{d+1} \gamma^a n_a$, where $\gamma_{d+1}$ is the
$d$-dimensional analog of the four-dimensional $\gamma_5$ matrix, so that it
obeys the relation $N^2 = 1$. Then we have\footnote{Here and below, we omit
the implicit label $s$ in the operator of reflection $N$.}
\begin{equation}
\psi_{(1)} = \pm N \psi_{(2)} \, , \quad \psi_{(2)} = \pm N \psi_{(1)} \, ,
\label{usual}
\end{equation}
where the sign ambiguity corresponds to the fact that the spinor
representation of the Lorentz group is double-valued.

Conditions (\ref{usual}) correspond to the continuity of the spinor field in
the space ${\mathcal M}$ thus indicating that the junction ${\mathcal B}$ in
this case is only imaginary, or nonphysical.  It will be shown very soon
that the junction conditions (\ref{usual}) may be obtained from the
variational principle starting from more general conditions, namely,
\begin{equation}
\psi_{(1)} = (\alpha + \beta N) \psi_{(2)} \, , \quad   \psi_{(2)} = (\alpha
+ \beta N) \psi_{(1)} \, , \label{general}
\end{equation}
where $\alpha$ and $\beta$ are some constants.  Thus, it appears reasonable
to impose the general junction conditions of the type (\ref{general}) in the
general case shown in Fig.~\ref{space-n}.

We proceed with the analysis of the situation shown in Fig.~\ref{space-n}
and impose the discussed general junction conditions at ${\mathcal B}$,
compatible with the symmetry of the problem:
\begin{equation}
\psi_{(s)} = \sum_{r \ne s} \left( \alpha + \beta N \right) \psi_{(r)} \, ,
\quad s = 1,\ldots,n \, , \label{general3}
\end{equation}
where $\alpha$ and $\beta$ are constants.

To facilitate the analysis, we can use the decomposition $\psi = \psi^+ +
\psi^-$ at ${\mathcal B}$, where
\begin{equation}
\psi^\pm = \frac12 (1 \pm N) \psi \, , \quad  N \psi^\pm = \pm \psi^\pm \, .
\label{decomp}
\end{equation}
Then the requirement that system (\ref{general3}) have nontrivial solutions
both for $\psi^+_{(s)}$ and for $\psi^-_{(s)}$, $s = 1,\ldots,n$, at the
junction leads to the equation
\begin{equation}
\left[(n - 1)z - 1\right]\, \left(z + 1\right)^{n - 1} =  0 \, ,
\label{cubic}
\end{equation}
which must be satisfied both by $z = \alpha + \beta$ and by $z = \alpha -
\beta$.

Two solutions of these equations for $\alpha$, $\beta$ have $\beta = 0$. In
this case, equation (\ref{cubic}) implies that either $\alpha = -1$, or
$\alpha = 1/(n - 1)$. It is easy to verify that the boundary term in the
variational principle (\ref{spinvar}) in both these cases will lead to the
additional condition $\psi_{(s)} = 0$, $s = 1, \ldots, n$, which, together
with the Dirac equation in the volume, will imply vanishing of $\psi$ all
over the space ${\mathcal M}$. Thus, the junction conditions with $\beta =
0$ should be rejected as leading to only trivial solutions.

Consider the remaining two solutions with nonzero $\beta$, namely,
\begin{equation}
\alpha = {2 - n \over 2(n - 1)} \, , \quad  \beta = \pm {n \over 2(n - 1)}
\, , \label{ab3}
\end{equation}
which differ in the sign of $\beta$.  By writing the fields on the junction
${\mathcal B}$ in the form (\ref{decomp}) and by using the property of the
reflection operator
\begin{equation}
N \gamma^a N n_a = - \gamma^a n_a \, , \label{prop}
\end{equation}
one easily can verify that both cases (\ref{ab3}) imply the identical
vanishing of the boundary term in variation (\ref{spinvar}) of the action.
Thus, both cases (\ref{ab3}) can describe the physical situation, and the
possible junction conditions at ${\mathcal B}$ can be summarized as
\begin{equation}
\psi_{(s)} = {1 \over 2(n - 1)} \sum_{r \ne s} \left[ (2 - n) \pm n N
\right] \psi_{(r)} \, , \quad s = 1, \ldots, n \, . \label{usual3}
\end{equation}
These junction conditions can be further transformed to the following
convenient form:
\begin{equation}
(1 \pm N) \psi_{(s)} = {2 \over n} \sum_r \psi_{(r)} \, , \quad s = 1,
\ldots, n \, . \label{usualn}
\end{equation}
In this form, the junction conditions are extended also to the case $n = 1$,
i.e., where ${\mathcal B}$ is the boundary of only one volume space; in this
case, the boundary conditions imply reflection from the boundary. In the
particular case of $n = 2$, we obtain precisely conditions (\ref{usual}),
which imply continuity of the Dirac field all over ${\mathcal M}$.  As in
the previous cases, we will refer to the junction conditions (\ref{usualn})
with the upper and lower sign as to the junction conditions of type A and B,
respectively, although in the case of the Dirac field, there is no
qualitative difference between them.

As in the scalar case, the junction conditions (\ref{usualn}) imply the sum
rule for the components of the conserved current $J^a = \overline \psi
\gamma^a \psi$ normal to the junction:
\begin{equation}
\sum_s J^a_{(s)} n^{(s)}_a = 0 \, .
\end{equation}

\section{Wave propagation through the junction} \label{wave}

In this section, we apply the equations obtained to the particular
interesting case of wave propagation in the space shown in
Fig.~\ref{space-n}.  We shall derive the laws of wave transmission through
and reflection from the junction ${\mathcal B}$.

First, consider the simple case of a scalar field.  Let, in the region
${\mathcal M}_1$, a wave that obeys the Klein--Gordon equation and
propagates towards the junction ${\mathcal B}$ be denoted by $\phi^{(+)}_1$.
We denote its value at ${\mathcal B}$ by $\phi_{\mathcal B}$ and its
derivative normal to the junction ${\mathcal B}$ by $\phi'_{\mathcal B}
\equiv n_{(1)}^a \nabla_a \phi^{(+)}_1$. Then the wave which we call the
reflected wave and denote by $\phi^{(-)}_1$ is constructed by imposing the
same values at the junction, $\phi^{(-)}_1 = \phi_{\mathcal B}$, and by
reversing the sign of the derivative normal to the junction ${\mathcal B}$:
$n_{(1)}^a \nabla_a \phi^{(-)}_1 = - \phi'_{\mathcal B}$. For example, in
the case of propagation in a flat spacetime ${\mathcal M}_1$ with the
surface ${\mathcal B}$ described by the equation $x_1 = 0$ in the natural
spacetime coordinates $t$, $x \equiv (x_1, x_2, \ldots, x_{d - 1})$, the
plane waves of this kind will be given, respectively, by $\phi^{(+)}_1 =
\exp (- i \omega t + i k \cdot x)$ and $\phi^{(-)}_1 = \exp (- i \omega t +
i k' \cdot x)$, where the wave vector $k'$ is obtained from $k$ by reversing
its $x_1$-component. The waves $\phi^{(-)}_s$, $s = 2,\ldots,n$, propagating
away from the junction, respectively, in the regions ${\mathcal M}_s$, $s =
2,\ldots,n$, are constructed by imposing the boundary conditions
$\phi^{(-)}_s = \phi_{\mathcal B}$, $n_{(s)}^a \nabla_a \phi^{(-)}_s = -
\phi'_{\mathcal B}$, $s = 2,\ldots,n$, at the junction ${\mathcal B}$.  We
will assume that solutions with the boundary conditions imposed exist
globally in ${\mathcal M}_s$, $s = 1,\ldots,n$, respectively.

We are looking for a solution that contains both waves falling towards
${\mathcal B}$ and reflected from ${\mathcal B}$ in the region ${\mathcal
M}_1$, but only waves propagating away from ${\mathcal B}$ (transmitted
waves) in the regions ${\mathcal M}_s$, $s = 2, \ldots, n$. Thus, we set
\begin{equation}
\phi_1 = \phi^{(+)}_1 + \rho \phi^{(-)}_1 \, , \quad  \phi_s = \tau_s
\phi^{(-)}_s \, , \quad  s = 2, \ldots, n \, ,
\end{equation}
where $\rho$ is the amplitude of wave reflection and $\tau_s$ are the
amplitudes of wave transmission to the spaces ${\mathcal M}_s$, $s = 2,
\ldots, n$, respectively.

To determine the amplitudes of reflection and transmission, we apply the
junction conditions obtained in Sec.~\ref{scalvec}.  In the case of the
junction conditions (\ref{scal1}), we obtain
\begin{equation}\left\{
\begin{array}{l}
1 + \rho = \tau_2 = \tau_3 = \cdots = \tau_n \, , \\ 1 - \rho - \sum_{s =
2}^n \tau_s = 0 \, ,
\end{array} \right. \Longrightarrow \left\{
\begin{array}{l}
\rho = (2 - n) / n \, , \\ \tau_2 = \tau_3 = \cdots = \tau_n = 2/n \, .
\end{array} \right. \label{upper}
\end{equation}
In the case of the junction conditions (\ref{scal2}), we get the solution
that differs from (\ref{upper}) only in the sign:
\begin{equation}\left\{
\begin{array}{l}
1 + \rho + \sum_{s = 2}^n \tau_s = 0 \, , \\ 1 - \rho = - \tau_2 = - \tau_3
= \cdots = - \tau_n \, ,
\end{array} \right. \Longrightarrow \left\{
\begin{array}{l}
\rho = (n - 2)/n \, , \\ \tau_2 = \tau_3 = \cdots = \tau_n = - 2/n \, .
\end{array} \right. \label{lower}
\end{equation}
We see that, in both cases, the same amount of energy [the fraction $(n
-2)^2 / n^2$] is reflected back to the space ${\mathcal M}_1$ and the same
equal amount of energy (the fraction $4 / n^2$) is transmitted to each of
the $n - 1$ spaces ${\mathcal M}_s$, $s = 2, \ldots, n$.

The results for the case of vector fields and for weak gravitational waves
are essentially the same.  For the vector field, we introduce the wave
$A_a^{(+)}$ propagating towards the junction ${\mathcal B}$ in the region
${\mathcal M}_1$ and construct the reflected wave $A_a^{(-)}$ by keeping the
component $A_a^{\parallel}$ tangent to ${\mathcal B}$ intact and by
reversing the sign of the value of $F^{ab}n_b$ at ${\mathcal B}$.  For a
weak gravitational wave, we introduce the similar field $\delta
g_{ab}^{(+)}$ and construct the corresponding reflected wave $\delta
g_{ab}^{(-)}$ by keeping the perturbation of the induced metric $\delta
h_{ab}$ at ${\mathcal B}$ intact and by reversing the sign of the
perturbation of the extrinsic curvature $\delta K_{ab}$ at ${\mathcal B}$.
Then, proceeding in precisely the same way as we did in the scalar case, we
obtain the same amplitudes of reflection and transmission. Again, the only
difference between cases A and B of Sec.~\ref{scalvec} for the vector field
is in the relative phases (signs of the amplitudes) with which the waves are
reflected and transmitted. The reflection and transmission amplitudes for
gravitational waves will be given by (\ref{upper}).

The case of propagation of the Dirac field is also considered quite
similarly to the scalar case.  We denote by $\psi^{(+)}_1$ the wave that
propagates towards the junction ${\mathcal B}$ in the region ${\mathcal
M}_1$, and by $\psi_{\mathcal B}$ we denote its value at ${\mathcal B}$.
Then the reflected wave $\psi^{(-)}_1$ is constructed by imposing the
reflection boundary condition $\psi^{(-)}_1 = N \psi_{\mathcal B}$ at
${\mathcal B}$ and by subsequently solving the Dirac equation in ${\mathcal
M}_1$. Similarly, the waves that propagate away from ${\mathcal B}$ in the
spaces ${\mathcal M}_s$, $s = 2, \ldots, n$, are constructed by imposing the
conditions $\psi^{(-)}_s = N \psi_{\mathcal B}$ at the junction ${\mathcal
B}$ and by subsequently solving the Dirac equation in ${\mathcal M}_s$.  We
assume that such solutions exist globally in ${\mathcal M}_s$, $s = 1,
\ldots, n$, as will be the case in a flat spacetime with a flat junction
hypersurface considered while discussing the scalar case. With waves thus
constructed, we set
\begin{equation}
\psi_1 = \psi^{(+)}_1 + \rho \psi^{(-)}_1\, , \quad  \psi_s = \tau_s
\psi^{(-)}_s \, , \quad  s = 2, \ldots, n \, .
\end{equation}
Again, $\rho$ is the coefficient of reflection, and $\tau_s$, $s = 2,
\ldots, n$, are the corresponding coefficients of transmission of waves.

In applying the junction conditions (\ref{usualn}), it is convenient to use
the decomposition $\psi = \psi^+ + \psi^-$ defined by (\ref{decomp}) at the
junction ${\mathcal B}$ and to write the junction conditions for the
$\psi^+$ and $\psi^-$ components separately.  With the upper sign in
(\ref{usualn}), we obtain precisely the set of equations (\ref{upper})
while, with the lower sign in (\ref{usualn}), we get precisely the system of
equations (\ref{lower}). Thus, we conclude that the laws of wave reflection
from and transmission through the junction ${\mathcal B}$ are similar for
all the spins considered.

\section{Green functions in flat spaces with flat miltivolume
junctions}\label{green}

In this section, we consider the simple case of flat spaces ${\mathcal M}_s$
with flat common timelike boundary ${\mathcal B}$ (see Fig.~\ref{space-n}).
We obtain the expressions for the Green functions in such a space.

In each space ${\mathcal M}_s$, $s = 1, \ldots, n$, we can choose natural
coordinates $x = \{x_1, \ldots, x_d\}$ in such a way that the junction
${\mathcal B}$ is the boundary surface $x_d = 0$ of the volume $x_d \ge 0$,
and the points with coordinates $\{x_1, \ldots, x_{d - 1}, 0\}$ in the
spaces ${\mathcal M}_s$ are naturally identified.  Let $G(x, x') \equiv
D(x_1 - x'_1, \ldots, x_d - x'_d)$ be any Green function (retarded,
advanced, causal, etc.\@) for the scalar field in the Minkowski space. Then,
the corresponding Green function in the space ${\mathcal M} = \cup_s
{\mathcal M}_s$ is easily constructed by the method of images.  We introduce
the function $\widetilde G(x, x') \equiv D(x_1 - x'_1, \ldots, x_d + x'_d)$.
Then the Green function $G_{\mathcal M}(x, x')$ in the space ${\mathcal M}$
is given by
\begin{equation}\label{greens}
G_{\mathcal M} (x, x') = \left\{ \begin{array}{rl} \displaystyle G (x, x')
\pm {2 - n \over n}\, \widetilde G (x, x') \, , \quad &\mbox{if \ $x \sim
x'$}\, , \medskip \\ \displaystyle \pm {2 \over n}\, \widetilde G (x, x') \,
, \quad &\mbox{if \ $x \not\sim x'$}\, ,
\end{array} \right.
\end{equation}
where $x \sim x'$ means that $x$ and $x'$ are in one component ${\mathcal
M}_s$, and $x \not\sim x'$ means that $x$ and $x'$ are in different
components ${\mathcal M}_s$.  The upper sign in (\ref{greens}) corresponds
to the junction conditions (\ref{scal1}), and the lower sign corresponds to
the junction conditions (\ref{scal2}).

Similar relations can be obtained for the Green functions of the vector and
Dirac fields.  For example, in the case of the Dirac field, the Green
function $G (x, x') \equiv D(x_1 - x'_1, \ldots, x_d - x'_d)$ is a matrix
with spinor indices.  Then we introduce the function $\widetilde G (x, x')
\equiv N D (x_1 - x'_1, \ldots, x_d + x'_d)$, where $N$ is the usual matrix
of reflection acting on the index corresponding to the argument $x'$, and,
using the junction conditions (\ref{usual3}), we arrive at the same form
(\ref{greens}) for the Green function.

Using expressions (\ref{greens}), one easily can obtain the renormalized
vacuum stress-energy tensor.  It is given by the derivatives of the Hadamard
function renormalized by subtracting the Hadamard function for the Minkowski
space (see \cite{BD}).  We obtain
\begin{equation}\label{stress-energy}
\left\langle T_{ab} \right\rangle_{\rm A} = {}- \left\langle T_{ab}
\right\rangle_{\rm B} = {2 - n \over n} \left\langle T_{ab}
\right\rangle_{\rm A}^{(n = 1)} \, ,
\end{equation}
where the labels `{\small A}' and `{\small B}' correspond to the junction
conditions of type A and B, respectively, and the expressions for
$\left\langle T_{ab} \right\rangle_{\rm A, B}^{(n = 1)}$ are standard and
can be found, e.g., in \cite{BD} and references therein.

The stress-energy tensor $\left\langle T_{ab} \right\rangle_{\rm A}^{(n =
1)}$ typically diverges as $x_d \to 0$.  For example, for a massless scalar
field, we have \cite{BD}
\begin{equation}
\left\langle T_{ab} \right\rangle_{\rm A}^{(n = 1)} = {1 \over 16 \pi^2
x_d^4}\, g_{ab}  \, .
\end{equation}
If stress-energy tensor of such form must be added to the matter side of the
Einstein equation in the volume, then its presence is inconsistent with the
assumption that the spacetime is flat. This constitutes a well-known problem
for curved spaces and, especially, spaces with boundaries (see \cite{BD}).
However, we may simply avoid this problem in the case under consideration by
requiring that exactly two copies of each field are present in the theory,
one with the junction conditions A, and another with the junction conditions
B. Then their contributions to the renormalized stress-energy tensor will
cancel each other, as is clear from Eq.~(\ref{stress-energy}). Additionally,
we must also consider quantum fluctuations of the metric that are expected
to result in an effective regularization of the stress-energy tensor in the
junction region. This will be the subject of the future investigations.

\section{Universe with multivolume junctions}\label{branching}

First, let us consider a universe with spatial three-dimensional topology as
shown in Fig.~\ref{branch}. Here, we have three spaces ${\mathcal M}_1$,
${\mathcal M}_2$, and ${\mathcal M}_3$ with topology of a three-dimensional
disk bounded by the common surface ${\mathcal B}$ that has topology of
two-sphere.

We assume that the topology described may arise in the context of the theory
of quantum creation of a closed universe on the background of a big mother
universe \cite{Fomin}. It is conceivable that the created baby universe does
not become spatially separated from the mother universe, but rather remains
glued with it over some common three-dimensional volume \cite{FSB}. Then
Fig.~\ref{branch} can be interpreted as showing the mother universe
${\mathcal M}_1$ and the baby universe ${\mathcal M}_2$ glued over the
volume ${\mathcal M}_3$.  All the three volume regions ${\mathcal M}_1$,
${\mathcal M}_2$, and ${\mathcal M}_3$ that have common boundary ${\mathcal
B}$ may evolve (expand or contract) preserving the topological configuration
shown.  One of the important physical questions in this situation is the
issue of the behavior of various physical fields, in particular, of the
metric, in this topology.

Consider the case where the metrics of the pieces ${\mathcal M}_1$,
${\mathcal M}_2$, and ${\mathcal M}_3$ are the usual
Friedmann--Robertson--Walker metrics given by the line element
\begin{equation}
ds^2 = dt^2 - a^2 (t) ds^2_3 \, , \label{frw}
\end{equation}
where
\begin{equation}
ds^2_3 = d \chi^2 + f^2 (\chi) d \Omega_2 \, ,
\end{equation}
\begin{equation}
f (\chi) = \left\{ \begin{array}{rll} \sin\chi \quad  &\mbox{for} \ &\kappa
= 1 \, , \\ \chi \quad &\mbox{for} &\kappa = 0 \, , \\ \sinh\chi \quad
&\mbox{for} &\kappa = - 1 \, , \end{array} \right.
\end{equation}
$d \Omega_2$ is the line element of the unit two-spherical geometry, and the
discrete parameter $\kappa$ indicates the type of the spatial geometry.  The
time coordinates $t_s$, the scale factors $a_s(t_s)$, the angles $\chi_s$,
and the functions $f_s (\chi_s)$ specified by the numbers $\kappa_s$ are to
be introduced for each space ${\mathcal M}_s$, $s = 1,2,3$, separately.

We consider the junction conditions (\ref{metric1}) for the metric field in
the absence of contribution to the right-hand side. Let the position of the
junction ${\mathcal B}$ be described by the function $\chi = \chi_* (t)$ in
the metric (\ref{frw}). The components of the extrinsic curvature of the
junction in the part of the space $\chi \le \chi_*$ are given by
\begin{equation}
K^t{}_t = - {d \over dt} \left[{a \dot \chi_* \over \sqrt{1 - \left(a \dot
\chi_*\right)^2}} \right] \, , \quad  K^i{}_j = { {}- \delta^i{}_j  \over
\sqrt{1 - \left(a \dot \chi_*\right)^2}} \left[\dot a \dot \chi_* + {f'
(\chi_*) \over a f (\chi_*)} \right] \, , \quad  i, j = 1,2 \, ,
\end{equation}
where overdot denotes the time derivative.

In general, possible motions of the junction in each of the three pieces of
the volume space will be determined by the junction conditions
(\ref{metric1}), and this is not an easy problem even in the symmetric case
that we are considering now. One special situation can be analyzed in the
case where the three spaces expand in a similar way so that their Hubble
parameters $H \equiv \dot a / a$ coincide as functions of time, which can be
chosen common to all three spaces. Then solutions exist for which $\dot
\chi_* \equiv 0$ in each of the spaces, i.e., the junction expands together
with the universe.  Introducing the radius $r = a f(\chi_*)$ of the
junction, we will have the following condition:
\begin{equation}
\sum_s \epsilon_s \sqrt{1 - \kappa_s \left(r / a_s\right)^2} = 0 \, ,
\label{frwbc}
\end{equation}
where $\epsilon_s = \mbox{sign\,$f_s'(\chi_{s*})$}$.  Note that $\epsilon_s
= 1$ for hyperbolic and flat spatial geometry ($\kappa_s = -1, 0$) while,
for spherical spatial geometry ($\kappa_s = 1$), $\epsilon_s = \pm 1$. Then,
with the topology shown in Fig.~\ref{branch}, one can conclude from
Eq.~(\ref{frwbc}) that at least {\em two\/} of the spaces ${\mathcal M}_s$
must have spherical spatial geometry.  Let these spaces be ${\mathcal M}_1$
and ${\mathcal M}_2$.  If, moreover, we suppose that $r \ll a_1, a_2$ and
$\epsilon_1 = \epsilon_2 = -1$ (the situation actually depicted in
Fig.~\ref{branch}) then it is necessary that the third space ${\mathcal
M}_3$ have hyperbolic spatial geometry and $r \approx \sqrt{3}\,a_3$.  To
avoid confusion, we stress that these conditions are valid only for scaling
solutions under consideration (identical Hubble parameters and $\dot \chi_*
\equiv 0$ in each of the spaces).  Also note that we have not analyzed the
matter content of such universes which is necessary to produce the desired
solutions.

The general laws of propagation of waves of various fields through the
junction ${\mathcal B}$ were described in the previous Sec.~\ref{wave}.  In
particular, as a wave reaches the junction ${\mathcal B}$ in the space
${\mathcal M}_1$, $1/3$ of its amplitude is reflected back to the space
${\mathcal M}_1$ while $2/3$ is transmitted to each of the spaces ${\mathcal
M}_2$ and ${\mathcal M}_3$.

Consider now the example of a universe whose spatial section has a boundary.
In this case, the boundary conditions (\ref{metric1}) with vanishing
right-hand side imply vanishing of the extrinsic curvature of the boundary.
Restricting analysis to the simple case of spherical boundary in a spatially
flat Friedmann--Robertson--Walker spacetime, we obtain the following
solution of the boundary conditions:
\begin{equation}\label{bound}
a(t) = t_0 \left({t \over t_0}\right)^q \, , \quad \chi(t) = \sqrt{q(q - 1)}
\left({t_0 \over t}\right)^{q - 1} \, , \quad q > 1 \, .
\end{equation}
The condition $q > 1$ is a consequence of the requirement that the boundary
be timelike, and it implies the power-law accelerating expansion of the
universe.  As follows from (\ref{bound}), the radius of the boundary $r
\equiv a \chi$ in the expanding universe evolves according to the law $r(t)
= t\,\sqrt{q(q - 1)}$.  Note that solution (\ref{bound}) can describe both a
space with an outer boundary (a disk) and a space with an inner boundary (a
space with a hole).

\section{Discussion}\label{discussion}

Spaces with topology as that shown in Fig.~\ref{space-n} naturally arise in
the theory of multivolume junctions in four-dimensional spacetimes, in the
theory of brane worlds, and in the superstring theory. It is therefore
important to study the possible junction conditions at the hypersurface
${\mathcal B}$ and their physical consequences.  In this paper, after
establishing the junction conditions, we studied the issue of field
propagation in spaces with the specified topology. It turns out that the
laws of wave transmission through and reflection from the junction are quite
similar for fields of all physical spins.

We considered the particular case of a multivolume junction in a
four-dimensional spacetime and presented a partial solution for the metric
with topology shown in Fig.~\ref{branch}.  The aim of the subsequent
investigations in this direction will be to investigate the case of a
universe with multivolume junctions in more detail and to study their
physical implications. One of the ideas is to identify regions of type
${\mathcal M}_3$ in Fig.~\ref{branch} with the observed voids (see, e.g.,
\cite{Peebles}) in the large-scale distribution of galaxies in the universe.

Boson string configurations of type shown in Fig.~\ref{space-n} with $n = 3$
were studied in \cite{string} with the natural junction conditions
(\ref{scal1}) for the target space coordinates on the string world sheet. It
would be interesting to study such configurations in the superstring theory
with the additional junction conditions (\ref{usualn}) for spinor fields.

In the case of integer spin, one may wish to view the junction conditions~A
[with $\alpha = 1/(n - 1)$] of Sec.~\ref{scalvec} as more physical than the
junction conditions~B (with $\alpha = -1$) since, in the first case, the
fields are continuous in the manifold ${\mathcal M}$ while, in the second
case, they are discontinuous at the surface ${\mathcal B}$. However, one
should not discard the junction conditions of type~B altogether before
studying them in greater detail.  This is supported by our example of flat
spaces with flat multivolume junctions considered in Sec.~\ref{green}, which
shows that the presence of fields with both types of junction conditions may
lead to cancellation of certain divergences in the vacuum stress-energy
tensor.

\section*{Acknowledgments}

The authors are grateful to V.~P.~Frolov and V.~D.~Gladush for useful discussion.
Yu.~S.\@ acknowledges support from the INTASgrant for project No.~2000-334.

\appendix
\section{Variation of the action for the metric}

Here, we derive the expression for the first variation of the action for
gravity (up to a multiplicative constant)
\begin{equation} \label{a-action}
S_g =  \int_{\mathcal M} R + 2 \int_{{\mathcal B}} K \, ,
\end{equation}
where ${\mathcal B}$ is the boundary of ${\mathcal M}$, $h_{ab} = g_{ab} +
n_a n_b$ is the metric induced on ${\mathcal B}$, $K_{ab} = h^c{}_a \nabla_c
n_b$ is the extrinsic curvature of ${\mathcal B}$, and $K = K_{ab}h^{ab}$ is
its trace. In contrast with the standard derivation, here we do not assume
that the variation of $g_{ab}$ vanishes at the boundary ${\mathcal B}$,
which is taken to be timelike.

We start from the standard expression (see, e.g., Appendix~E of \cite{Wald},
but note that we work with the opposite metric signature and use the {\em
inner\/} spacelike normal $n^a$)
\begin{equation} \label{var1}
\delta \left( \int_{\mathcal M} R \right) = \int_{\mathcal M} G_{ab}\,
\delta g^{ab} + \int_{\mathcal M} \nabla^a v_a \, ,
\end{equation}
where
\begin{equation}
v_a = \nabla^b \left(\delta g_{ab} \right) - g^{cd} \nabla_a \left(\delta
g_{cd} \right) \, .
\end{equation}
The second integral in (\ref{var1}) can be transformed with the use of the
Stokes theorem as
\begin{equation} \label{a-vn}
\int_{\mathcal M} \nabla^a v_a = \int_{\mathcal B} v_a n^a \, ,
\end{equation}
where
\begin{equation}
v_a n^a = n^a g^{bc} \left[ \nabla_c \left( \delta g_{ab} \right) - \nabla_a
\left( \delta g_{bc} \right) \right] = n^a h^{bc} \left[ \nabla_c \left(
\delta g_{ab} \right) - \nabla_a \left( \delta g_{bc} \right) \right] \, .
\end{equation}

 Then we have
\begin{equation} \label{a-deltaK}
\delta K = \delta \left( h^a{}_b \nabla_a n^b \right) = \delta h^a{}_b
\nabla_a n^b + h^a{}_b (\delta C)^b{}_{ac} n^c + h^a{}_b \nabla_a \delta n^b
\, ,
\end{equation}
where
\begin{equation}
(\delta C)^b{}_{ac} = {1 \over 2}\, g^{bd} \left[ \nabla_a \left(\delta
g_{cd}\right) + \nabla_c \left(\delta g_{ad}\right) - \nabla_d \left(\delta
g_{ac}\right) \right] \, .
\end{equation}
The first term in the right-hand side of (\ref{a-deltaK}) is identically
zero. Indeed, we have $\delta n_a = n_a n_b \delta n^b$, so that
\begin{equation}\begin{array}{r}
\delta h^a{}_b \nabla_a n^b =  \left( \delta n^a n_b + n^a \delta n_b
\right) \nabla_a n^b = \left( \delta n^a + n^a n_c \delta n^c \right) n_b
\nabla_a n^b
\smallskip \\ = \delta n^c h^a{}_c n_b \nabla_a n^b = \delta n^c n_b K_c{}^b = 0
\, .
\end{array}
\end{equation}

 Thus, variation of the second term of (\ref{a-action}) is
\begin{equation} \label{a-varK}
\delta \left( 2 \int_{\mathcal B} K \right) = \int_{\mathcal B} \left[ n^c
h^{ab} \nabla_c \left( \delta g_{ab} \right) + 2 h^a{}_b \nabla_a \delta n^b
- K h_{ab} \delta h^{ab} \right] \, ,
\end{equation}
where the last term in the square brackets stems from the variation of the
volume element $\sqrt{|h|}\, d^{d - 1} x$ in the integral over ${\mathcal
B}$.

The total boundary term in the variation of action (\ref{a-action}) is given
by the sum of (\ref{a-vn}) and (\ref{a-varK}) with the result
\begin{equation}
(\mbox{Boundary term}) = \int_{\mathcal B} \left[ n^a h^{bc} \nabla_c \left(
\delta g_{ab} \right) + 2 h^a{}_b \nabla_a \delta n^b - K h_{ab} \delta
h^{ab} \right] \, .
\end{equation}

 We transform the first term in the integrand of the last expression:
\begin{equation}
n^a h^{bc} \nabla_c \left( \delta g_{ab} \right) = h^{bc} \nabla_c \left(
n^a \delta g_{ab} \right) - h^{bc} \nabla_c n^a \delta g_{ab} = h^{bc}
\nabla_c \left( n^a \delta g_{ab} \right) + K_{ab} \delta h^{ab} \, .
\end{equation}
Then
\begin{equation} \label{a-boundary}
(\mbox{Boundary term}) = \int_{\mathcal B} \left[ h^{bc} \nabla_c \left( n^a
\delta g_{ab} \right) + 2 h^a{}_b \nabla_a \delta n^b \right] +
\int_{\mathcal B} \left(K_{ab} - K h_{ab}\right) \delta h^{ab} \, .
\end{equation}

 Now we show that the integrand of the first integral in (\ref{a-boundary})
is a divergence, so that this integral vanishes for variations of $g_{ab}$
with compact support in ${\mathcal B}$.  Indeed,
\begin{equation} \label{a-divergence}
\begin{array}{r}
h^{bc} \nabla_c \left( n^a \delta g_{ab} \right) + 2 h^a{}_b \nabla_a \delta
n^b = h^{bc} \nabla_c \left( \delta n_b - g_{ab} \delta n^a \right) + 2
h^a{}_b \nabla_a \delta n^b \smallskip \\ = h^a{}_b \nabla_a \left( g^{bc}
\delta n_c + \delta n^b \right) = h^a{}_b \nabla_a \left(h^b{}_c \delta n^c
\right) = D_b \left(h^b{}_c \delta n^c \right) \, ,
\end{array}
\end{equation}
where $D_a$ is the (unique) derivative on ${\mathcal B}$ associated with the
induced metric $h_{ab}$, and the last equality in (\ref{a-divergence}) is
valid by virtue of Lemma~10.2.1 of \cite{Wald}.

 As a final result, we obtain
\begin{equation}
\delta S_g = \int_{\mathcal M} G_{ab}\, \delta g^{ab} + \int_{\mathcal B}
\left(K_{ab} - K h_{ab}\right) \delta h^{ab}\, .
\end{equation}

\end{document}